\documentstyle{mn}

\input epsf

\begin{document}

\def\Lya{Ly$\alpha\ $}
\def\Lyb{Ly$\beta\ $}
\def\Lyg{Ly$\gamma\ $}
\def\Lyd{Ly$\delta\ $}
\def\Lye{Ly$\epsilon\ $}
\def\LCDM{$\Lambda$CDM\ }
\def\HI{\hbox{H~$\rm \scriptstyle I\ $}}
\def\HII{\hbox{H~$\rm \scriptstyle II\ $}}
\def\DI{\hbox{D~$\rm \scriptstyle I\ $}}
\def\HeI{\hbox{He~$\rm \scriptstyle I\ $}}
\def\HeII{\hbox{He~$\rm \scriptstyle II\ $}}
\def\HeIII{\hbox{He~$\rm \scriptstyle III\ $}}
\def\CIV{\hbox{C~$\rm \scriptstyle IV\ $}}
\def\NHI{N_{\rm HI}}
\def\NHeII{N_{\rm HeII}}
\def\cm2{\,{\rm cm$^{-2}$}\,}
\def\kms{\,{\rm km\,s$^{-1}$}\,}
\def\kmsmpc{\,{\rm km\,s$^{-1}$\,Mpc$^{-1}$}\,}
\def\hmpc{\,h^{-1}{\rm \,Mpc}\,}
\def\mpch{\,h{\rm \,Mpc}^{-1}\,}
\def\ev{\,{\rm eV\ }}
\def\kel{\,{\rm K\ }}
\def\intunits{\,{\rm ergs\,s^{-1}\,cm^{-2}\,Hz^{-1}\,sr^{-1}}}
\def\ltsima{$\; \buildrel < \over \sim \;$}
\def\lsim{\lower.5ex\hbox{\ltsima}}
\def\gtsima{$\; \buildrel > \over \sim \;$}
\def\gsim{\lower.5ex\hbox{\gtsima}}
\def\etal{{ et~al.~}}
\def\aj{AJ}
\def\apj{ApJ}
\def\apjs{ApJS}
\def\mnras{MNRAS}

\journal{Preprint-00}

\title{Particle-Mesh Simulations of the Lyman-Alpha Forest}

\author[A. Meiksin and M. White]{Avery Meiksin${}^{1}$, Martin White${}^{2}$ \\
${}^1$Institute for Astronomy, University of Edinburgh,
Blackford Hill, Edinburgh\ EH9\ 3HJ, UK \\
${}^2$Astronomy Department, Harvard University
60 Garden Street, Cambridge, MA 02138 , USA}

\pubyear{2000}

\maketitle

\begin{abstract}
Numerical hydrodynamical simulations have proven a successful means of
reproducing many of the statistical properties of the Lyman-Alpha forest as
measured in high redshift quasar spectra. Pseudo-hydrodynamical methods based
only on simulating the dark matter component have been claimed to yield a
comparable level of success. We investigate the degree to which two
pseudo-methods, with and without allowing for a pseudo-gas pressure, are able
to match the predictions of fully hydrodynamical plus dark matter simulations. 
We also address the requirements for convergence
to the statistics of the spectra and the inferred properties of the Lyman-Alpha
forest as a function of resolution and box size. Generally we find it is
possible to reach agreement with full hydrodynamic simulations at the 10\%
level in the cumulative distributions of the flux and absorption line parameter
statistics for readily achievable particle and grid numbers, but difficult to
do much better.
\end{abstract}

\section{Introduction}
\label{sec:introduction}

Numerical simulations of structure formation in the universe incorporating
hydrodynamics in Cold Dark Matter (CDM) dominated cosmologies have proven
very successful in reproducing the statistical properties of the \Lya forest
as measured in high redshift Quasi-Stellar Object (QSO) spectra
(Cen \etal 1994; Zhang, Anninos \& Norman 1995; Hernquist \etal 1996;
Zhang \etal 1997; Bond \& Wadsley 1997; Theuns, Leonard \& Efstathiou 1998a).
The high level of statistical agreement suggests that
the models are capturing the essential physical nature of the absorbing
structures (Meiksin \etal 2000). Much of the structure of the
Intergalactic Medium (IGM) may be understood as a consequence of the
spatial coherence in the statistical properties of the density fluctuations
of the dark matter alone in a ``cosmic web'', as emphasized by Bond \& Wadsley
and supported by the slow comoving evolution of the integalactic filaments
found by Zhang \etal (1998). Gnedin \& Hui (1998) claim that in fact it is
unnecessary to incorporate full hydrodynamics in the
simulations to reproduce the properties of the \Lya forest. Gnedin \& Hui
introduce instead a pseudo-hydrodynamical scheme into an N-body calculation,
reporting comparable results in the distributions of flux and \HI column
density to full hydrodynamical $+$ N-body simulations for a standard CDM model.
Pseudo-hydrodynamical methods based on PM alone have been used by
Petitjean, M\"ucket \& Kates (1995) and Croft \etal (1998).

Because of the large computational overhead in incorporating
proper hydrodynamics into an N-body code, considerable benefit would
be obtained if the hydrodynamics computations could be avoided. It would
then be possible to perform a wide range of cosmological models at only
moderate computational expense, as well as test systematics on the
properties of the forest:\ the effect of box size, resolution, and
even physical effects like variations in the character of the
background UV radiation field that ionizes the gas. It would be of
additional theoretical interest to infer the degree to which the properties
of the forest are hydrodynamical in origin, as opposed to being solely a
consequence of the clustering of dark matter.

To these ends, we perform simulations of the \Lya forest for a variety
of cosmological models using two approaches:\ a pure gravitational computation
using a Particle-Mesh N-body code (PM simulations), and a pseudo-hydrodynamical
technique, based on the method of Gnedin \& Hui, incorporating an effective gas
pressure force computed from the local dark matter density (HPM simulations).

\section{The Models and Simulations}
\label{sec:models}

\begin{figure}
\begin{center}
\leavevmode \epsfxsize=3.3in \epsfbox{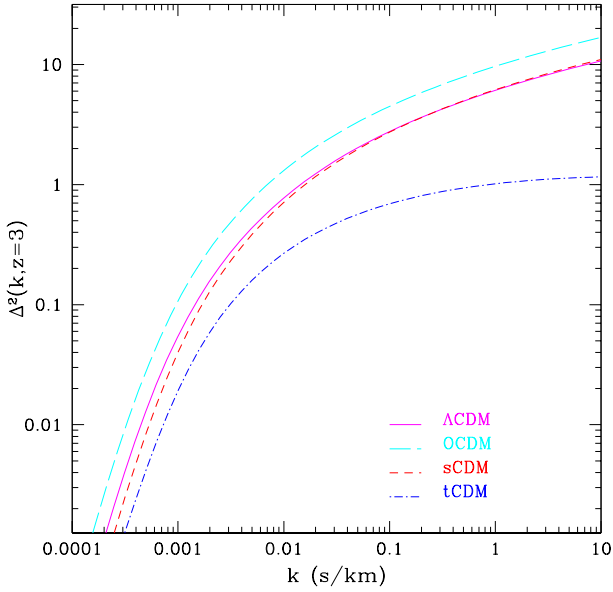}
\end{center}
\caption{Non-dimensional linear power spectra $\Delta^2(k)=k^3P(k)/(2\pi^2)$,
where $P(k)$ is the power spectrum, for the four cosmological models
examined. Shown at $z=3$.}
\label{fig:PS}
\end{figure}

We consider two sets of models. The first is based on simulations in a
flat CDM model with a cosmological constant (\LCDM) using both the PM
and HPM codes. These are performed in order to test convergence of the
results for differing box sizes, particle numbers, and grid resolutions.
A second set is based on 4 CDM models in Machacek \etal (2000), who performed
computations using full hydrodynamics. We duplicate their models
in order to compare our resulting statistics for the absorption
features with theirs. The models investigated, in additional to the
\LCDM model, are:\ a standard critical-density flat CDM model without a
cosmological constant (SCDM), an open CDM model (OCDM),
and the standard CDM model but with the power spectrum of the density
perturbations tilted (tCDM) to match the normalization on large scales
as determined from the COBE measurements of the Cosmic Microwave
Background (Bunn \& White 1997). The model parameters are shown in
Table \ref{tab:model_par}. As in Machacek \etal, the initial
data were generated using the BBKS
transfer function (Bardeen \etal 1986) to compute the starting
redshifts and the unconstrained initial particle positions and
velocity perturbations appropriate for each model. The non-dimensional
linear power spectra $\Delta^2(k)$ are shown at $z=3$ in Figure~\ref{fig:PS}.

To account for the degree of ionization of the hydrogen, it is also
necessary to include a background UV radiation field. We essentially
circumvent this by fixing the mean optical depth, as defined by
$\bar\tau_\alpha=-\log<{\rm flux}>$, at $\bar\tau_\alpha=0.30$ at $z=3$,
as in Machacek \etal

The evolution is computed using the PM code described in detail in
Meiksin, White \& Peacock~\cite{MWP} and White~\cite{whi99}.
The computational volume is chosen to be periodic with side length 1,
and we take as our time coordinate the log of the scale factor, $\log a$,
where $a\equiv (1+z)^{-1}$.  Velocities are measured in units of the expansion
velocity across the box $aHL_{\rm box}$, where $H$ is the Hubble parameter
fixed by the Friedmann equation
\begin{equation}
  H^2 \equiv H_0^2\left( {\Omega_{\rm mat}\over a^3} + \Omega_\Lambda
  + {\Omega_K\over a^2} \right)
\end{equation}
in terms of the Hubble constant $H_0$ and the densities in matter, cosmological
constant and curvature, in units of the critical density.  The density is
defined from the particle positions by assigning particles to a regular
cartesian grid using the Cloud-In-Cell (CIC) scheme (Hockney \&
Eastwood~\cite{HocEas}).  The Fourier Transform is taken and the force computed
using the kernel $\vec{k}/k^2$.
We use a second order leap frog method to integrate the equations.
The relevant positions are predicted at a half time step and used to calculate
the accelerations which modify the velocities.  The time step is dynamically
chosen as a small fraction of the free-fall time with a maximum
size of $\Delta\log a=3\%$.

The initial conditions were created by displacing the particles from a uniform
grid using the Zel'dovich approximation.

Our implementation of the HPM code differs slightly from that in Gnedin \& Hui.
Firstly we use 2 sets of particles, one of which feels only the gravitational
force (the `dark matter') and one of which additionally feels a pseudo-pressure
force (the `gas').  The gravity is computed as before.  To compute the `hydro'
forces we bin the `gas' particles onto a grid using CIC as for the gravity.
We compute the temperature using the equation of state
\begin{equation}
  T = T_0\left( {\rho\over\bar{\rho}} \right)^{\gamma-1}
\end{equation}
where $T_0$ and $\gamma$ are obtained from Hui \& Gnedin (1997). The enthalpy
per particle
\begin{equation}
  \varphi = {\gamma\over\gamma-1} T
\end{equation}
is then computed, Fourier transformed and smoothed by a gaussian of half a
grid cell (to supress high frequency noise).  The `hydro' force is then
calculated from the inverse Fourier transform of $i\vec{k}\widetilde{\varphi}$
and added to the gravitational force for the `gas' particles.

In practice, $T_0$ will depend on the reionization history of the IGM. In
particular, if \HeII is reionized (to \HeIII) late, lower values of $T_0$
will result prior to \HeII reionization and higher values in rarified gas
after (Meiksin 1994). The effect of reionization on the thermal properties
of the gas will only be known once the nature of the ionizing sources has been
established, and may eventually require the incorporation of radiative
transfer into the simulations.

We consider a variety of particle number and grid zone combinations, as well
as several box sizes. The PM simulations are summarised in
Table \ref{tab:sim_par}. For the HPM simulations, two combinations are
used:\ $(N_p,N_g) = (128, 256)$ and $(256, 512)$. For HPM the dark matter and
the gas are represented by an identical number of particles,
each equal to $N_p^3$.

Given a set of final particle positions and velocities, we compute the spectra
as follows. First the density and density-weighted line-of-sight velocity
are computed on a grid (using CIC interpolation as above) and Gaussian smoothed
using FFT techniques.  This forms the fundamental data set.
A grid of sightlines is drawn through the box, parallel to the box sides and
along each sightline a new 1D grid of density and velocity are obtained from
the fundamental grid using CIC interpolation.  The position in velocity units
and the temperature are computed on this 1D grid using the cosmological model
and equation (2) for the equation of state.

Using this 1D grid we then integrate in real space to find $\tau(u)$ at a
given velocity $u$.  Specifically we define
\begin{equation}
  \tau(u) = A \int dx \left[ {\rho(x)\over\bar{\rho}} \right]^2 T(x)^{-0.7}
b^{-1} e^{-(u-u_0)^2/b^2}
\end{equation}
where $u_0=xaHL_{\rm box}+v_{\rm los}$ and $b=\sqrt{2k_B T/ m_{\rm H}}$ is the
Doppler parameter, where $m_{\rm H}$ is the mass of a hydrogen atom.
The flux at velocity $u$ is $\exp[-\tau]$. The integration variable $x$
indicates the distance along the box in terms of the expansion velocity across
the box. The constant $A$ is then iteratively adjusted to obtain a predefined
$\bar\tau_\alpha=-\log\left\langle\exp(-\tau)\right\rangle$.

We generate sample spectra from the simulated data along random
lines--of--sight through the computational boxes over the length of each box.
In order to make a fair comparison with the data in M00, both sets of spectra
are rebinned to a constant pixel resolution of $\lambda/\Delta\lambda=74000$
and gaussian smoothed to mimic a spectral resolution of 8\kms.
These parameters were chosen to approximate what may be achieved using
the Keck HIRES.

There are several measurements of the \Lya forest which may be used as a
basis for comparison. The most fundamental is simply the distribution of
flux per pixel in the synthesized spectra. The fluctuations in the flux
on different velocity scales provide additional constraints, which may
be related both to the density fluctuation power spectrum and to the thermal
widths of the lines. We quantify the fluctuations in three ways. We compute the
power spectrum $P_F(k)$ of the flux, an analog of the density power spectrum.
(For the $P_F(k)$ estimates we use the raw spectra from the PM and HPM runs.)
The widths of the spectral features may be directly characterised using
wavelets, which quantify the changes in a given spectrum on being smoothed
from one velocity resolution to another (Meiksin 2000). A multiscale analysis
based on the Daubechies wavelets effectively performs a weighted
smoothing of the spectrum over successive doublings
of the pixel width. We use a Daubechies wavelet of order 20. Lastly,
the most traditional method to quantify the flux fluctuations is by
decomposing each spectrum into a set of Voigt line profiles. We utilise
AutoVP (Dav\'e et al. 1997) to perform such an analysis.
Each sample spectrum produces an average of on the order of 900 lines per
unit redshift. Typically $(3-6)\times10^4$ lines are analysed per model.
The absorption lines are characterised by their \HI column densities
and Doppler parameters. The \HI column densities reflect both the
density and physical thickness of the structures which give rise to
the absorption features. The Doppler parameters reflect the degree
of line broadening due both to the thermal motion of atoms and any
internal peculiar velocities within the structures, and may provide
probes of the temperature and velocity history of the universe at high
redshifts.

To ensure our results are not affected by sample variance, we performed two
independent simulations of some of the models. We found convergence on the
statistical properties could be achieved with a single realization provided
the statistics were averaged over a sufficient number of lines-of-sight. We
found 3072 line-of-sight (each one box side in length) was adequate for the PM
simulations, and 1024 for HPM.
\section{Convergence Tests}
\label{sec:tests}

Simulations of the \Lya forest are restricted by two size constraints.
High spatial resolution is required to resolve the structure of the
absorbers, while a large box size is required both to capture the
large-scale power which will affect the velocity widths of the absorption
lines as well as to obtain a fair sample of the universe. Although a small
box may be adequate to resolve the line structure, if the box is too small
the fluctuations corresponding to the scale of the box will become nonlinear,
in which case the results are no longer representative of the cosmological
model simulated.

Resolution effects were previously studied using full hydrodynamics by Theuns
\etal (1998b) and Bryan \etal (1999). Using P$^3$M-SPH, Theuns \etal
found for SCDM that a (comoving) box size of at least 5.5~Mpc using $64^3$ SPH
particles and $64^3$ dark matter particles is required to converge on
$\bar\tau_\alpha$ and the Doppler parameter distributions at $z=3$. This
corresponds to a mean proper interparticle separation of 21~kpc, or about half
this in structures with an
overdensity of $10$, typical of the filaments with \HI column densities of
$10^{14}-10^{15}$\cm2 (Zhang \etal 1998). Bryan \etal found that a box size of
at least 9.6~Mpc was needed to reach convergence on the Doppler parameter
distribution at $z=3$ using a PM-PPM code (Kronus) with $128^3$ grid cells,
corresponding to a proper resolution of 19~kpc.

Neither group was able to maintain a resolution of 10~kpc while increasing
box sizes beyond 5~Mpc due to memory limitations, so that the contribution of
larger scale power at high resolution was left unexplored. We also note that
SCDM has less large-scale power than more viable cosmological models like
\LCDM, so that the resolution limitations are least restrictive for
SCDM. It is possible to achieve much higher resolution using PM or HPM.
In this section we present the results of a series of tests of the convergence
of the PM and HPM simulations for \LCDM as the box size, particle number, and
number of grid zones are varied. The simulation results are all shown at $z=3$.

\subsection{Flux distribution}
\label{sec:test_flux}

\begin{figure}
\begin{center}
\leavevmode \epsfxsize=3.3in \epsfbox{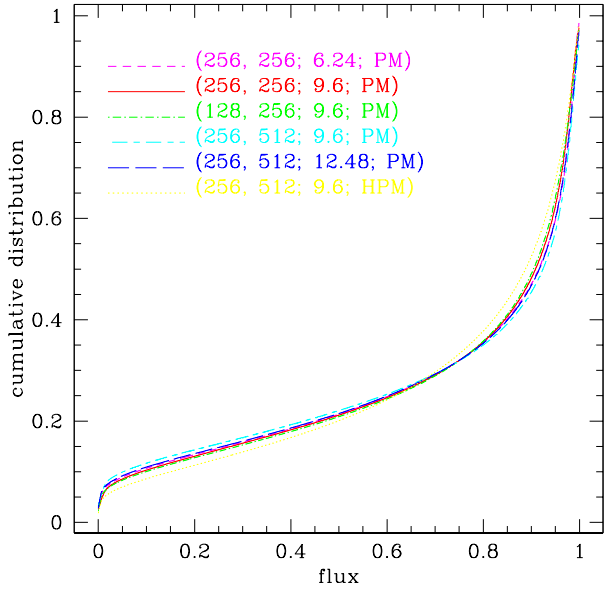}
\end{center}
\caption{Comparison of cumulative distributions of flux per pixel for the
$\Lambda$CDM simulations for three different resolutions and three box sizes.
The notation for each model is $({\rm N_p, N_g; box size; method})$, where the
number of dark matter particles is $N_p^3$, the number of grid zones is
$N_g^3$, and the boxsize is in units of $h^{-1}\,{\rm Mpc}$. For the HPM
simulations, the number of gas and dark matter particles are each $N_p^3$.
The distributions diverge at high flux values (small optical depths),
suggesting small structures have not been resolved at the lower resolutions.}
\label{fig:LCDM_flux_rescomp}
\end{figure}

\begin{figure}
\begin{center}
\leavevmode \epsfxsize=3.3in \epsfbox{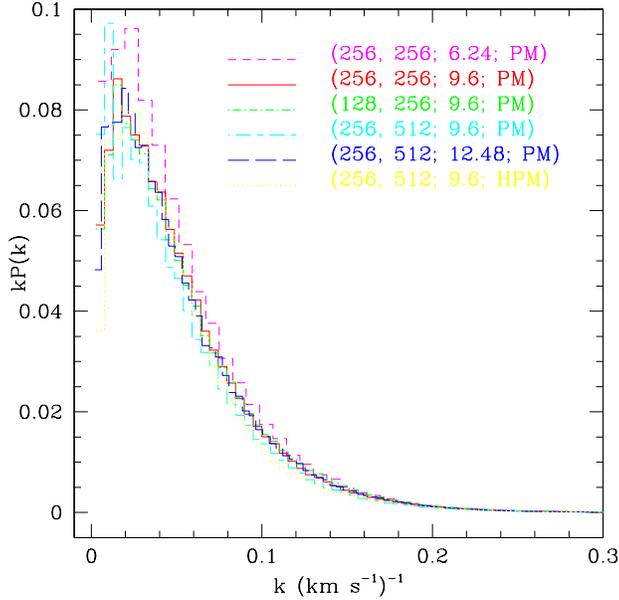}
\end{center}
\caption{Comparison of flux $P_F(k)$ for the $\Lambda$CDM simulations for
three grid resolutions and three box sizes. The notation for each model is
as in Figure 2. A higher amount of power is produced in the smallest box
simulation compared with the others. The HPM model suppresses power at
high $k$.}
\label{fig:LCDM_fluxpk_rescomp}
\end{figure}

A comparison of the cumulative distributions of flux per pixel for several
resolutions and box sizes is shown in Figure~\ref{fig:LCDM_flux_rescomp}.
The PM-generated distributions agree at low flux values (high optical depths),
but diverge for fluxes above 0.8, with differences in the cumulative
distributions as large as 0.05. As the resolution of the $9.6\hmpc$ box is
increased to $(N_p, N_g) = (256, 512)$, the amount of fine-scale structure
continues to increase. The HPM-generated flux distributions differ markedly
from the PM results, showing a smaller amount of fine-scale structure.
The corresponding flux power spectra $P_F(k)$ are show in
Figure~\ref{fig:LCDM_fluxpk_rescomp}. Again, as the grid resolution is
increased in the $9.6\hmpc$ box, the amount of large-scale power continues to
increase, while there is a decrease in the amount of power at shorter scales
($k>0.04\mpch$). The $9.6\hmpc$ HPM simulations show a deficit in power
compared with the $9.6\hmpc$ PM simulations for $k>0.08\mpch$, suggesting
that the gas pressure forces have suppressed power at short scales.

\subsection{Wavelet decomposition}
\label{sec:test_wavelets}

\begin{figure}
\begin{center}
\leavevmode \epsfxsize=3.3in \epsfbox{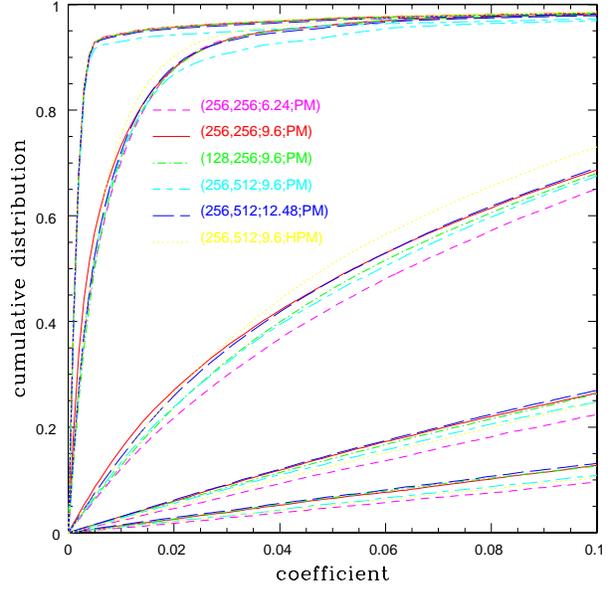}
\end{center}
\caption{Comparison of wavelet coefficient distributions for the $\Lambda$CDM
simulations for 3 resolutions and 3 box sizes.}
\label{fig:LCDM_wavec_rescomp}
\end{figure}

The cumulative distributions of wavelet coefficients are shown in
Figure~\ref{fig:LCDM_wavec_rescomp}. The groups of curve from left to
right correspond to the respective velocity scales $4-8$\kms,
$8-16$\kms, $16-32$\kms, $32-64$\kms, and $64-128$\kms. The small
coefficients at the lowest scales show that no features are present on
these scales. The physically most interesting scales are represented
by the $16-32$\kms curves, corresponding to typical velocity widths of
the absorption lines. The largest differences between the simulations
are found on these scales. The convergence for the $9.6\hmpc$ box
simulations is non-monotonic with $N_p$ and $N_g$, with the
$(N_p, N_g)=(128, 256)$
and $(256, 512)$ simulations agreeing well, but not with the $(256,256)$
simulation. The latter however agrees closely with the $(256,512)$
$12.48\hmpc$ simulation, suggesting that convergence has been
reached with box size. The largest coefficients are found for the
smallest box ($6.24\hmpc$) simulation, indicating a greater amount of
structure in the spectra on these velocity scales than found for the
larger boxes. By contrast, the HPM simulation shows the least amount
of structure on these scales.

\subsection{Line parameter decomposition}
\label{sec:test_lines}

\begin{figure}
\begin{center}
\leavevmode \epsfxsize=3.3in \epsfbox{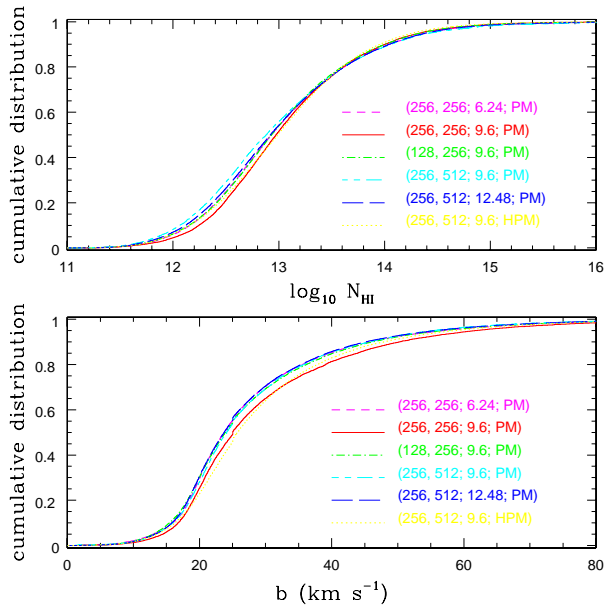}
\end{center}
\caption{Comparison of cumulative distributions of absorption line parameters
for the $\Lambda$CDM simulations for 3 resolutions and 3 box sizes.}
\label{fig:LCDM_avp_rescomp}
\end{figure}

A comparison of the cumulative distributions of $\NHI$ and $b$ determined
by AutoVP is shown in Figure~\ref{fig:LCDM_avp_rescomp} for PM simulations
for 3 different pairs of $N_p$ and $N_g$ and 3 box sizes.
As was found for the wavelet coefficient distributions, the behaviours for
both the $\NHI$ and b distributions are non-monotonic. The
$(N_p, N_g)=(128, 256)$ and $(256, 512)$ $9.6h^{-1}$~Mpc simulations agree
well in the b-distribution, but disagree with the $(N_p, N_g)=(256, 256)$
simulations. It would appear that AutoVP is picking up unreal features
resulting from noise in the $N_p=128$ simulations that decrease in going
to $N_p=256$. Increasing the grid resolution, holding $N_p=256$ fixed, however,
appears to recover real features that correspond statistically to the
previously unresolved fluctuations in the $(N_p, N_g)=(128, 256)$ simulations.
Since the same trend is found for the wavelet coefficients, it seems unlikely
that the non-monotonic behaviour is an artefact of the line finding and
fitting procedures used by AutoVP. There are no obvious systematic
trends with box size, the results otherwise appearing to have converged
to better than 5\%.

Also shown are the results for HPM for the $9.6\hmpc$ box and
$(N_p, N_g)=(256, 512)$.
Both the column density distribution and the Doppler parameter
distribution agree well with the $N_p=N_g=256$ $9.6\hmpc$ PM simulation,
demonstrating that incorporating the pseudo-pressure forces in the HPM
simulation has very little effect on the resulting absorption line parameter
distributions for this model. We note that the HPM simulation tends to produce
distributions shifted somewhat to higher column density and Doppler parameter
compared with the analogous $(N_p, N_g) = (256, 512)$ PM simulation.

\begin{figure}
\begin{center}
\leavevmode \epsfxsize=3.3in \epsfbox{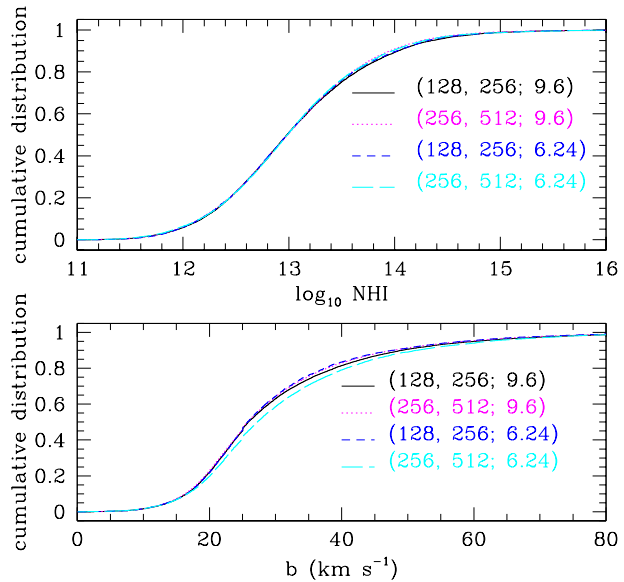}
\end{center}
\caption{Comparison of cumulative distributions of absorption line
parameters  for the HPM $\Lambda$CDM simulations for 2 grid resolutions
and 2 box sizes, showing good convergence.}
\label{fig:LCDM_avp_HPM_rescomp}
\end{figure}

A comparison of the results for HPM for different box sizes and particle/grid
numbers is shown in Figure~\ref{fig:LCDM_avp_HPM_rescomp}. The $\NHI$
distributions are well-converged. The $b$-distributions would have appeared
to converge, except the distributions shift toward larger $b$-values in the
highest resolution, small box simulation. Curiously, this is opposite the
trends reported in the resolution studies of Theuns \etal and Bryan \etal,
who found that the $b$-values tended to decrease both with decreasing box
size and increasing resolution, although at lower resolutions compared with
our simulations. The trends found in the full hydrodynamics simulations are
physically sensible:\ a larger box adds longer wavelength power to the
simulations, which may increase the peculiar velocities of the dark matter
and the gas, while higher particle and grid numbers permit higher resolution
of density fluctuations and shocks, which may result in a narrowing of
absorption features. Possibly the opposing trend found here is a consequence
of not allowing for shocks. For a typical sound speed of $\sim10$\kms and
peculiar velocities in the gas of $\sim100$\kms, shocks may be expected to
play an important role in the velocity structure of the lines.

\section{Comparison with Hydrodynamical Simulations}
\label{sec:hydro}

In this section we compare the results of the PM and HPM simulations against
the Kronus simulation results of Machacek \etal (2000) (M00) for a variety of
cosmological models. The simulations were run for the same cosmological
models as in M00, matching the box sizes and the numbers of hydrodynamical
cells and grid points for the gravitational force calculations.

\subsection{Flux distribution}
\label{sec:comp_flux}

\begin{figure}
\begin{center}
\leavevmode \epsfxsize=3.3in \epsfbox{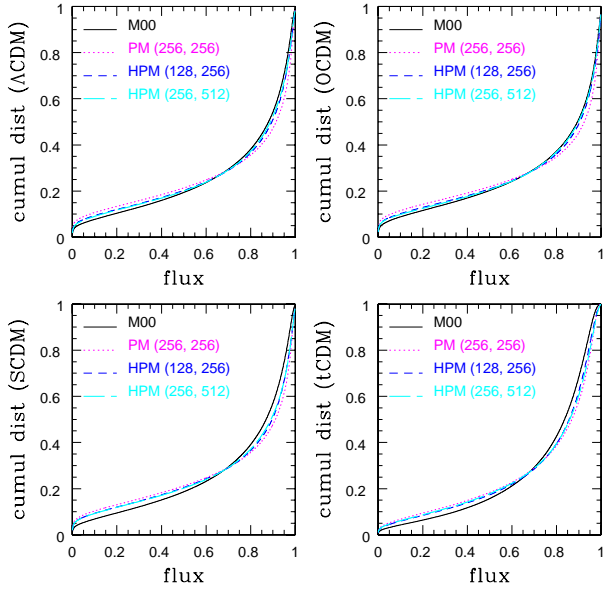}
\end{center}
\caption{Comparison of cumulative distributions of flux per pixel for four
cosmological models corresponding to the box sizes in M00. The particle and
grid numbers are shown as $(N_p, N_g)$.}

\label{fig:LCDM_flux_PM_NCSA}
\end{figure}

The distributions of flux for the 4 cosmological models considered are
shown in Figure~\ref{fig:LCDM_flux_PM_NCSA}. Excellent agreement is found
between HPM and the full hydrodynamical simulations (M00), with the
agreement least good for SCDM. The agreement with PM is not as good, but
the cumulative distributions still agree to within 10\%. The agreement
between both sets of $N_p$ and $N_g$ pairs for the HPM simulations shows
that the results have converged at the relevant box sizes.

\subsection{Wavelet decomposition}
\label{sec:comp_wavelets}

\begin{figure}
\begin{center}
\leavevmode \epsfxsize=3.3in \epsfbox{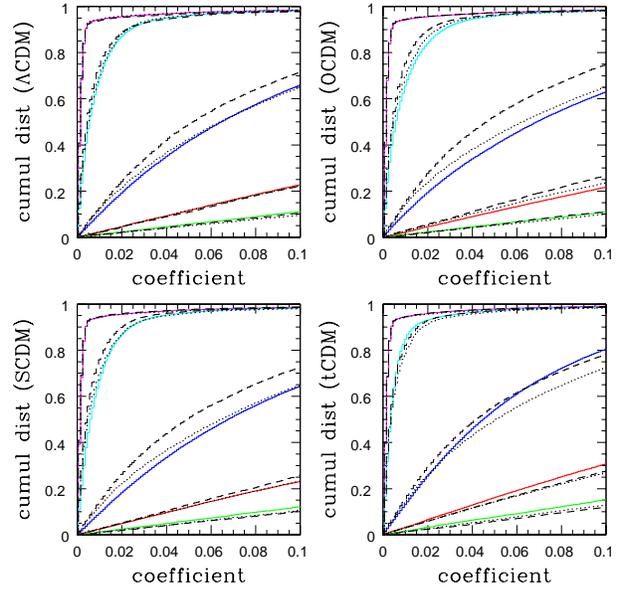}
\end{center}
\caption{Comparison of wavelet coefficient distributions for PM, HPM, and the
simulations of M00 (solid lines). The PM results are shown for $N_p=N_g=256$
(dotted lines) and the HPM results for $N_p=128$, $N_g=256$ (dashed lines).
The groups of coefficient distributions are ordered as in Figure 4.}
\label{fig:LCDM_wavec_PM_NCSA}
\end{figure}

The cumulative distributions for the wavelet coefficients are shown in
Figure~\ref{fig:LCDM_wavec_PM_NCSA}. Much better agreement is generally found
between the PM simulations and the full hydrodynamical simulations than is
found for HPM. The agreement is even poorer for the HPM simulations with
$(N_p, N_g)=(256, 512)$ in all cases. The smaller values for the coefficients
in the HPM spectrum indicates that the artificial hydrodynamics in HPM
suppresses the full amount of velocity structure in the spectra compared with
that found in the full hydrodynamics calculations.

\subsection{Line parameter decomposition}
\label{sec:comp_lines}

\begin{figure}
\begin{center}
\leavevmode \epsfxsize=3.3in \epsfbox{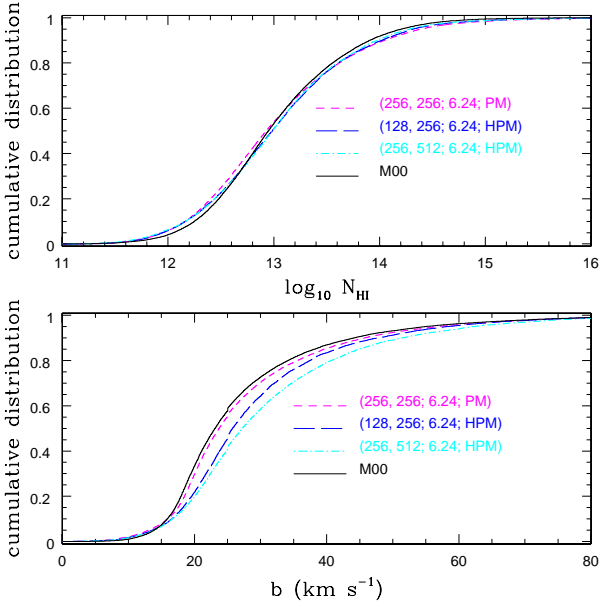}
\end{center}
\caption{Comparison of line parameter distributions for PM, HPM, and the
simulations of M00 for \LCDM.}
\label{fig:LCDM_avp_PM_NCSA}
\end{figure}

\begin{figure}
\begin{center}
\leavevmode \epsfxsize=3.3in \epsfbox{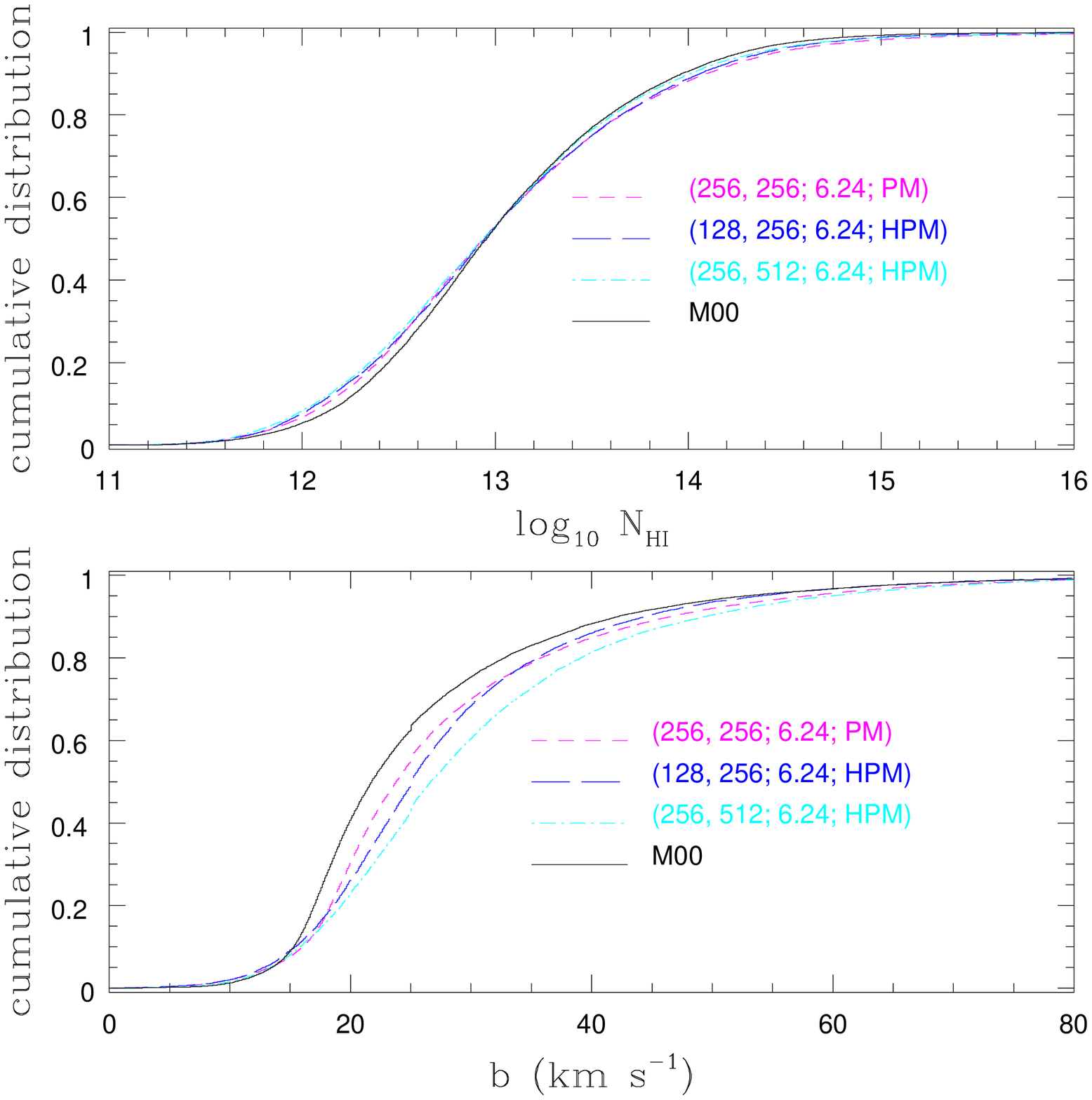}
\end{center}
\caption{Comparison of line parameter distributions for PM, HPM, and the
simulations of M00 for OCDM.}
\label{fig:OCDM_avp_PM_NCSA}
\end{figure}

\begin{figure}
\begin{center}
\leavevmode \epsfxsize=3.3in \epsfbox{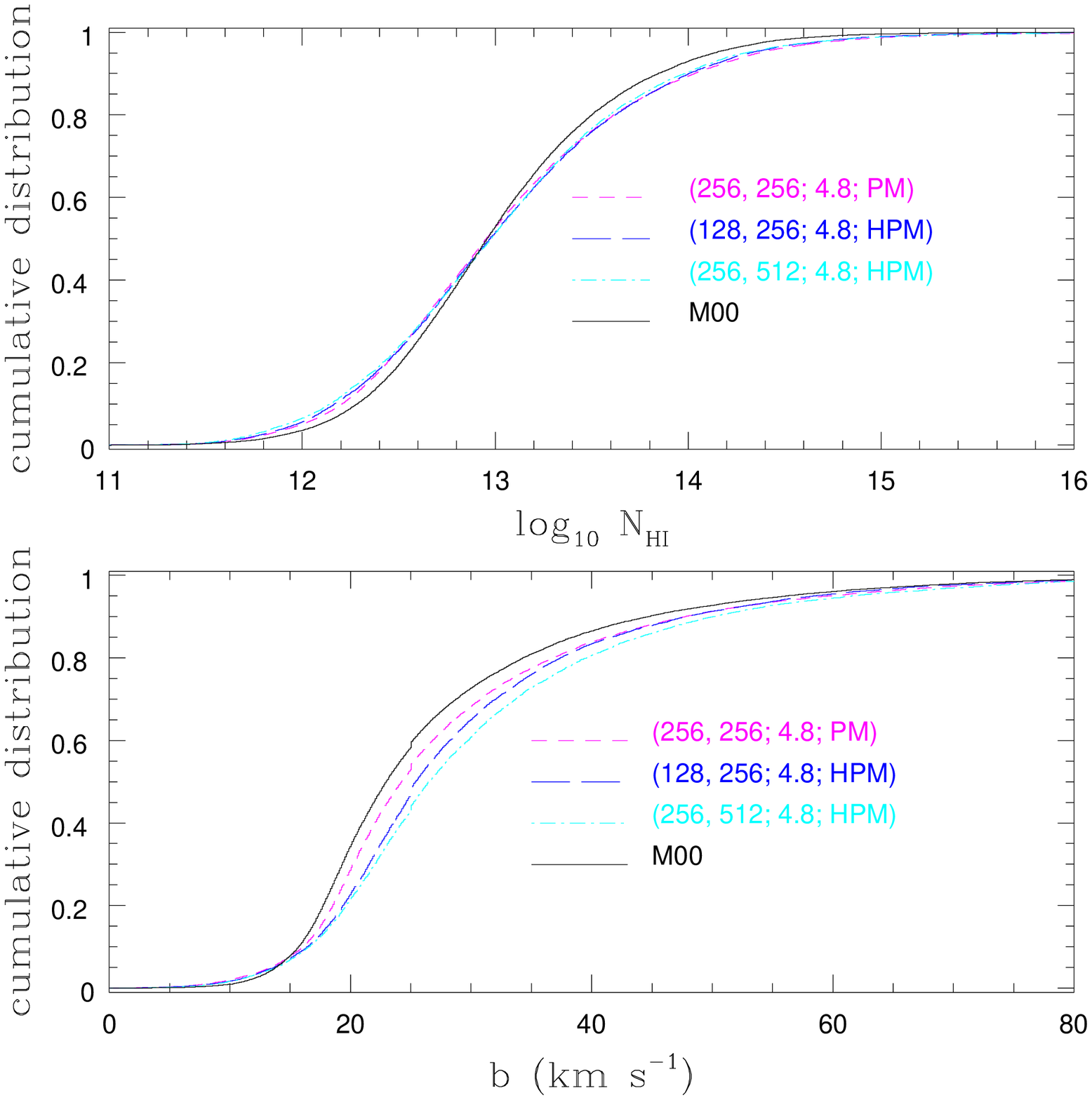}
\end{center}
\caption{Comparison of line parameter distributions for PM, HPM, and the
simulations of M00 for SCDM.}
\label{fig:SCDM_avp_PM_NCSA}
\end{figure}

\begin{figure}
\begin{center}
\leavevmode \epsfxsize=3.3in \epsfbox{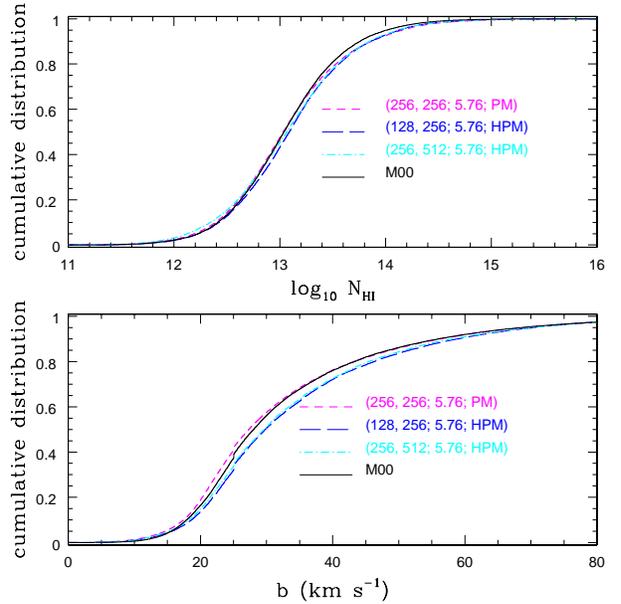}
\end{center}
\caption{Comparison of line parameter distributions for PM, HPM, and the
simulations of M00 for tCDM.}
\label{fig:TCDM_avp_PM_NCSA}
\end{figure}

The results of the Voigt line parameter analyses using AutoVP are shown
in Figures~\ref{fig:LCDM_avp_PM_NCSA} to \ref{fig:TCDM_avp_PM_NCSA}. Both the
PM and HPM results agree reasonably well with the results of M00 for the \HI
column density distributions. The best agreement for the Doppler parameter
distributions is found for the PM simulations:\ the HPM simulations produce
features that are too broad. While the PM and HPM simulations agree with the
results of M00 to $5-10$\% on the $\NHI$ distributions, the agreement between
the Doppler parameter distributions is worse:\ $\sim10$\% for PM and
$\sim20$\% for HPM. The disagreement between the Doppler parameter
distributions tends to increase with the magnitude of the power spectra
at $z=3$.

\section{Summary}
\label{sec:summary}

We have performed several simulations of the \Lya forest using
different background cosmological models, numerical codes and
grid resolutions. Convergence to the distributions of flux, wavelet
coefficients, and absorption line parameters tends to be non-monotonic
for the PM simulations as the number of particles and grid zones is
increased, although agreements in the cumulative distributions to within
10\% are achieved.

A comparison with full hydrodynamical calculations
shows that there is little advantage in HPM over PM, except possibly
for the flux distributions. The HPM simulations generally result in
broader absorption lines than PM. Both the PM and HPM simulations tend
to yield broader lines than found in the full hydrodynamics calculations.
This may be a consequence of the neglect of shocks in the PM and HPM
simulations, which may sharpen the velocity structure of the absorption lines.

The simulations show that the principal spectral properties of the \Lya forest
may be qualitatively reproduced from the dark matter structures alone, to a
remarkable accuracy of 10\% in the cumulative distributions of flux,
wavelet coefficients, \HI column density, and Doppler parameter. Tests of
the models, however, require agreement at the 5\% level per spectrum or better
at the resolution of the Keck HIRES (Meiksin \etal 2000), so that ultimately
full hydrodynamical calculations will be required to confront models for the
structure and physical properties of the \Lya forest with direct observations.
In order to match the observed spectra, it may ultimately be necessary to
incorporate radiative transfer into the simulations both to allow for spatial
fluctuations in the UV ionizing background and to reproduce the correct gas
temperatures (and line widths), particularly if helium is completely
photoionized at only moderate redshifts. Several other physical effects may
also be important, such as heating sources in addition to QSOs or galactic
feedback like winds. These effects may have sufficient impact on the
statistics of the \Lya forest that 10\% accuracy is adequate to constrain them.
Our results suggest that PM simulations alone may provide a reliable tool for
investigating some of these issues without the considerable overhead of solving
the full set of hydrodynamics equations.

\bigskip
\section*{acknowledgments}

The authors thank G. Bryan for permission to use the results of his Kronos
simulations, and R. Dav\'e for permission to use AutoVP.
M.~White was supported by the US National Science Foundation and a
Sloan Fellowship.
Parts of this work were done on the Origin2000 system at the National
Center for Supercomputing Applications, University of Illinois,
Urbana-Champaign.

\begin{table}
\centerline{\begin{tabular}{|c|c|c|c|c|c|c|} \hline
Model & $\Omega_0$ & $\Omega_\Lambda$   & $\Omega_b$ & $h$   & $n$        & $\sigma_{8h^{-1}}$ \\
\hline 
\hline
\LCDM & 0.4 & 0.6 & 0.0355 & 0.65 & 1    & 1.0 \\ \hline
OCDM  & 0.4 & 0   & 0.0355 & 0.65 & 1    & 1.0 \\ \hline
SCDM  & 1   & 0   & 0.06   & 0.5  & 1    & 0.7 \\ \hline
tCDM  & 1   & 0   & 0.07   & 0.6  & 0.81 & 0.5 \\ \hline
\end{tabular}}
\caption{Parameters for the cosmological models.
$\Omega_0$ is the total mass density parameter, $\Omega_\Lambda$ the 
cosmological constant density parameter, $\Omega_b$ the baryonic mass fraction,
$h=H_0/ 100$\kmsmpc, where $H_0$ is the Hubble constant at $z=0$,
$n$ the slope of the primordial density perturbation power spectrum, and
$\sigma_{8h^{-1}}$ the fluctuation normalization in a sphere of
radius $8h^{-1}$ Mpc.
}
\label{tab:model_par}
\end{table}

\begin{table}
\centerline{\begin{tabular}{|c|c|c|c|c|} \hline
Model & $L$ ($h^{-1}\,$Mpc) & $\Delta^2(2\pi/L)$ & $N_p$ & $N_g$ \\
\hline 
\hline
\LCDM & 6.24 & $0.64$ & 256 & 256 \\ \hline
\LCDM & 9.6 & $0.45$ & 128 & 256 \\ \hline
\LCDM & 9.6 & $0.45$ & 256 & 256 \\ \hline
\LCDM & 9.6 & $0.45$ & 256 & 512 \\ \hline
\LCDM & 12.48 & $0.35$ & 256 & 512 \\ \hline
OCDM  & 6.24 & $0.98$ & 256 & 256 \\ \hline
SCDM  & 4.8 & $0.50$ & 256 & 256\\ \hline
tCDM  & 5.76 & $0.19$ & 256 & 256 \\ \hline
\end{tabular}}
\caption{Physical parameters for the PM simulations.
$L$ is the comoving box size in units of $h^{-1}\,$Mpc.
Also shown is the dimensionless
linear power spectrum $\Delta^2(2\pi/L)$ at $z=3$. The number of dark matter
particles used is $N_p^3$ and the number of grid zones for the force
calculation is $N_g^3$.
}
\label{tab:sim_par}
\end{table}


\end{document}